\documentclass[prb,aps,12pt,a4paper,showpacs]{revtex4}
\usepackage{epsfig}
\begin{document}

\title{A Hamiltonian functional for the linearized Einstein vacuum field equations}
\author{R. Rosas-Rodr{\'\i}guez}

\address{Instituto de F{\'\i}sica, Universidad Aut\'onoma de Puebla, Apdo. Postal J-48,
72570, Puebla, Pue., M\'exico}

%\ead{rrosas@sirio.ifuap.buap.mx}

%\date{\today}

\begin{abstract}
By considering the Einstein vacuum field equations linearized
about the Minkowski metric, the evolution equations for the
gauge-invariant quantities characterizing the gravitational field
are written in a Hamiltonian form by using a conserved functional
as Hamiltonian; this Hamiltonian is not the analog of the energy
of the field. A Poisson bracket between functionals of the field,
compatible with the constraints satisfied by the field variables,
is obtained. The generator of spatial translations associated with
such bracket is also obtained.
\end{abstract}
\maketitle
%\pacs{PACS numbers:}

\section{Introduction}
\,\,\,\ As we know, the Lagrangian and the Hamiltonian formalisms
employed in the study of mechanical systems with a finite number
of degrees of freedom can be applied in the case of
infinite-dimensional systems. The Hamiltonian formulation is
usually obtained from the Lagrangian formulation by means of the
Legendre transformation, but in the case of fields this canonical
procedure presents difficulties since not always the momentum
densities are independent of the field variables, which is usually
mended by the introduction of constraints. Nevertheless, it is
possible to avoid these complications and give a Hamiltonian
formulation for a given continuous system, without making
reference to the Lagrangian formulation, if its evolution
equations can be written in the form
\begin{equation}
\dot{\phi} _{a}= D_{ab} \frac{\delta H}{\delta\phi_{b}},
\label{eq1}
\end{equation}
where the field variables $\phi _{a}(a=1,2,...,n)$ represent the
state of the system, $H$ is a suitable functional of the $\phi
_{a}$, $\delta H/\delta\phi _{b}$ is the functional derivative of
$H$ with respect to $\phi _{b}$, and the $D_{ab}$ are, in general,
operators that must satisfy certain conditions that allow the
definition of a Poisson bracket between functionals of the $\phi
_{a}$ (see, e.g. Refs. [1] and [2]). Here and henceforth a dot
denotes partial differentiation with respect to the time and there
is summation over repeated indices.

In Ref. [3] the evolution equations for the gravitational field,
given by the Einstein vacuum field equations linearized about the
Minkowski metric, are written in a Hamiltonian form (\ref{eq1}) in
terms of gauge-invariant quantities only by using an analog of the
energy of the electromagnetic field as Hamiltonian.

In this paper we propose a conserved functional of the field as a
new Hamiltonian and then we find a Hamiltonian structure for the
linearized Einstein theory. This conserved functional is found in
the Maxwell theory [5,6] and we use its analog in the linearized
Einstein theory as Hamiltonian. By contrast with the Hamiltonian
structure found in Ref. [3], which involves integral operators, in
the present case the $D_{ab}$ turn out to be constants.

In the next section the linearized Einstein vacuum field equations
are written in a noncovariant manner, emphasizing their analogy
with Maxwell's equations, in Sect.\ 3 we propose the Hamiltonian,
and in Sect.\ 4 such equations are written in a Hamiltonian form.
A Poisson bracket, compatible with the constraints imposed by the
field equations, is obtained and it is shown that it yields the
expected relations between the Hamiltonian or the momentum (which
we give also) and any functional of the field. Throughout this
paper Greek indices run from 0 to 3 and Latin indices, $i,j,...,$
from 1 to 3. Greek indices are raised and lowered by means of the
Minkowski metric. Repeated Latin lower indices are to be summed as
though a Kronecker delta $\delta^{ij}$ were present
$a_{i}b_{i}=\delta^{ij}a_{i}b_{j}.$

\section{The linearized Einstein vacuum field equations}
\,\,\,\
In the linearized Einstein theory it is assumed that, in a
suitable coordinate system, the metric of the space-time can be
written in the form
\begin{equation}
g_{\alpha \beta}= \eta_{\alpha \beta} + h_{\alpha \beta}
\label{eq2}
\end{equation}
where $h_{\alpha \beta}$  represents a small deviation of the
metric $g_{\alpha \beta}$  from the Minkowski metric
\begin{equation}
( \eta_{\alpha \beta}  ) \equiv {\rm diag} ( -1, 1, 1,
1).\label{eq3}
\end{equation}
The coordinate system in which expression (\ref{eq2}) applies is
not defined uniquely; under any ``infinitesimal coordinate
transformation", $ x'^{\alpha} = x^{\alpha}+\xi^{\alpha}$, the
metric has again the form (\ref{eq2}) with
\begin{equation}
h'_{\alpha \beta} =  h_{\alpha \beta} - \partial_{\alpha}
\xi_{\beta} - \partial_{\beta} \xi_{\alpha}, \label{eq4}
\end{equation}
where $\partial_{\alpha} \equiv \partial / \partial x^{\alpha} $.
The tensor field
\begin{equation}
K_{\alpha \beta \gamma \delta} \equiv \frac{1}{2} \left\{
\partial_{\alpha}
\partial_{\gamma} h_{\delta \beta} - \partial_{\beta} \partial_{\gamma}
h_{\delta \alpha} + \partial_{\beta} \partial_{\delta} h_{\gamma
\alpha} -
\partial_{\alpha} \partial_{\delta} h_{\gamma \beta} \right\}, \label{eq5}
\end{equation}
which is the curvature tensor corresponding to the metric
$g_{\alpha \beta}= \eta_{\alpha \beta} + h_{\alpha \beta}$ to
first order in $h_{\alpha \beta}$, is invariant under the gauge
transformations (\ref{eq4}). From its definition it is clear that
$K_{\alpha \beta \gamma \delta}$ possesses the symmetries
\begin{equation}
K_{\alpha \beta \gamma \delta} = - K_{\beta \alpha \gamma \delta}
= -K_{\alpha \beta \delta \gamma} = K_{\gamma \delta \alpha
\beta}, \label{eq6}
\end{equation}
\begin{equation}
K_{\alpha \beta \gamma \delta} + K_{\alpha \delta \beta \gamma} +
K_{\alpha \gamma \delta \beta} = 0, \label{eq7}
\end{equation}
and that it also satisfies the identities
\begin{equation}
\partial_{\alpha} K_{\beta \gamma \delta \varepsilon} + \partial_{\varepsilon}
K_{\beta \gamma \alpha \delta} + \partial_{\delta} K_{\beta \gamma
\varepsilon \alpha} = 0. \label{eq8}
\end{equation}
Conversely, Eqs. (\ref{eq6}-\ref{eq8}) imply that, locally,
$K_{\alpha \beta \gamma \delta}$ has the form (\ref{eq5}) where
$h_{\alpha \beta}$ is some symmetric tensor field defined up to
the transformations (\ref{eq4}).

In terms of the right dual $K^{\ast}_{\alpha \beta \gamma \delta}$
of $K_{\alpha \beta \gamma \delta}$ defined by
\begin{equation}
K^{\ast}_{\alpha \beta \gamma \delta} \equiv \frac{1}{2} K_{\alpha
\beta}{}^{\rho \sigma}\epsilon_{\rho \sigma\gamma\delta}
\label{eq9}
\end{equation}
where $\epsilon_{\alpha \beta \gamma \delta}$ is completely
antisymmetric with $\epsilon_{0123}=1$, Eqs. (\ref{eq7}) and
(\ref{eq8}) are equivalent to
\begin{equation} K^{\ast}_{\alpha\beta\gamma}{}^{\beta} = 0 \label{eq10}
\end{equation}
and
\begin{equation} \partial^{\gamma}K^{\ast}_{\alpha\beta\gamma\delta} = 0, \label{eq11}
\end{equation}
respectively. From Eqs. (\ref{eq6}) and (\ref{eq8}) it follows
that
\begin{equation}
K^{\ast}_{\alpha \beta \gamma \delta} = - K^{\ast}_{\beta \alpha
\gamma \delta} = -K^{\ast}_{\alpha \beta \delta \gamma}
\label{eq12}
\end{equation}
which are analogous to the first two equalities in (\ref{eq6}).
Nevertheless, in general, $K^{\ast}_{\alpha \beta \gamma \delta}$
does not possess all the symmetries of $K_{\alpha \beta \gamma
\delta}$   [Eqs. (\ref{eq6}, \ref{eq7})]. In fact, from the
definition (\ref{eq8}) one obtains that
\begin{equation}
K^{\ast}_{\alpha \beta \gamma \delta} - K^{\ast}_{\gamma
\delta\alpha\beta} = \frac{1}{2} \left\{ \epsilon_{\alpha \beta
\delta\rho}K_{\gamma}{}^{\rho}+\epsilon_{\beta\alpha\gamma\rho}
K_{\delta}{}^{\rho}+\epsilon_{\gamma\beta
\delta\rho}K_{\alpha}{}^{\rho}+\epsilon_{\delta\alpha\gamma
\rho}K_{\beta}{}^{\rho} \right\},\label{eq13}
\end{equation}
where the tensor field
\begin{equation}
K_{\alpha \beta} \equiv  K_{\alpha \gamma \beta}{}^{\gamma},
\label{eq14}
\end{equation}
which is symmetric as a consequence of Eqs. (\ref{eq6}), has been
introduced. Similarly, one finds that
\begin{equation}
K^{\ast}_{\alpha \beta \gamma \delta} + K^{\ast}_{\alpha \delta
\beta \gamma} + K^{\ast}_{\alpha \gamma \delta \beta} =
-\epsilon_{\beta \gamma\delta\rho}K_{\alpha}{}^{\rho}.
\label{eq15}
\end{equation}
On the other hand, from the identities (\ref{eq8}) it follows that
\begin{equation}
\partial^{\alpha} K_{\beta \alpha \delta \varepsilon} = -
\partial_{\varepsilon} K_{\beta \alpha}{}^{\alpha}{}_{\delta} -
\partial_{\delta} K_{\beta \alpha \varepsilon}{}^{\alpha} =
\partial_{\varepsilon} K_{\beta \delta} - \partial_{\delta}K_{\beta
\varepsilon}, \label{eq16}
\end{equation}

The linearized Einstein field equations are
\begin{equation}
K_{\alpha \beta} - \frac{1}{2} \eta_{\alpha \beta}
K_{\gamma}{}^{\gamma} = - \frac{8 \pi G}{c^{4}} T_{\alpha \beta}
\label{eq17}
\end{equation}
or, equivalently
\begin{equation}
K_{\alpha \beta} = - \frac{8 \pi G}{c^{4}}\left(T_{\alpha \beta} -
\frac{1}{2} \eta_{\alpha \beta}T_{\gamma}{}^{\gamma}\right),
\label{eq18}
\end{equation}
where $T_{\alpha \beta}$ is the energy-momentum tensor of the
matter to first order in  $h_{\alpha \beta}$. Therefore, the
linearized Einstein vacuum field equations are $K_{\alpha
\beta}=0$ and from Eqs. (\ref{eq6},\ref{eq7}) and
(\ref{eq9}-\ref{eq15}) one sees that $K^{\ast}_{\alpha \beta
\gamma \delta}$ satisfies the same relations as $K_{\alpha \beta
\gamma \delta}$ if and only if $K_{\alpha \beta}=0$ .

In what follows it will be assumed that the conditions $K_{\alpha
\beta}=0$ hold. This implies that all the components $K_{\alpha
\beta\gamma\delta}$  can be expressed in terms of the fields
$E_{ij}$ and $B_{ij}$ defined by
\begin{equation}
E_{ij} \equiv  K_{0i0j},\,\,\,\,\, B_{ij} \equiv -K^{\ast}_{0i0j},
\label{eq19}
\end{equation}
where the minus sign is introduced for later convenience. As a
consequence of Eqs. (\ref{eq6}), (\ref{eq10}), (\ref{eq13}) and
(\ref{eq14}), the fields $E_{ij}$ and $B_{ij}$ are symmetric and
have vanishing traces, hence each of them has five independent
components. Equations (\ref{eq11}) and (\ref{eq16}) amount to
\begin{equation}
\partial_{i}E_{ij}=0,\,\,\,\,\,\,\, \partial_{i}B_{ij}=0
\label{eq20}
\end{equation}
and
\begin{equation}
\frac{1}{c}\dot{E}_{ij}=\epsilon_{ikm}\partial_{k}B_{mj},\,\,\,\,\,\,\,
\frac{1}{c}\dot{B}_{ij}=-\epsilon_{ikm}\partial_{k}E_{mj},
\label{eq21}
\end{equation}
where $\epsilon_{ijk}$ is completely antisymmetric with
$\epsilon_{123}=1$, which are analogous to the source-free Maxwell
equations. It is easy to see that, due to Eqs. (\ref{eq20}) and to
the fact that $E_{ij}$ and $B_{ij}$ have vanishing trace, the
right-hand sides of Eqs. (\ref{eq21}) are symmetric in the indices
$i$ and $j$. Equations (\ref{eq20}), which do not involve time
derivatives, can be regarded as constraints on the fields $E_{ij}$
and $B_{ij}$.

\section{Hamiltonian functional}
\,\,\,\ As it was already mentioned in the introduction, in Ref.
[3] a Hamiltonian structure for the linearized Einstein theory is
found by using as Hamiltonian density the analog of the energy of
the electromagnetic field, $H= \int
\kappa(E_{ij}E_{ij}+B_{ij}B_{ij})dv/2$ (where $\kappa$ is a
constant), having to introduce ad hoc modifications in order to
get consistency with the constraints imposed by the field
variables. This Hamiltonian structure involves integral operators.

In the Maxwell theory we can see easily that the functional
$H=\int
c\epsilon_{ijk}(E_{i}\partial_{j}E_{k}+B_{i}\partial_{j}B_{k})dv/2$
is a conserved functional [5, 6, 7], and it can be used as
Hamiltonian in that theory. By analogy with the electromagnetic
field one can introduce the conserved functional
\begin{equation}
H=\int\mathcal{H}dv =\frac{c}{2}\int
\epsilon_{ikm}\left(E_{ij}\partial_{k}E_{mj}+B_{ij}\partial_{k}B_{mj}
\right)dv \label{eq22}
\end{equation}
as Hamiltonian in the linearized Einstein theory (we suppose that
the fields vanish at infinity).

We point out that a functional $F$ is a conserved functional if
and only if $dF/dt=0$. Therefore, to check if a functional of the
field is a conserved functional one needs to use the evolution
equations for the field only without having to choose a particular
Hamiltonian $H$ and its corresponding Hamiltonian structure
$D_{ab}$. Of course, if one makes a choice of the pair $(H,
D_{ab})$, then one can also use this knowledge to check it.

\section{Hamiltonian structure}
\,\,\,\ Equations (\ref{eq21}) can be written in the Hamiltonian
form
\begin{equation}
\dot{E}_{ij}= D_{ijkm} \frac{\delta H}{\delta
B_{km}},\,\,\,\,\,\,\, \dot{B}_{ij}= -D_{ijkm} \frac{\delta
H}{\delta E_{km}} \label{eq23}
\end{equation}
where
\begin{equation}
D_{ijkm}=\frac{1}{2}\left(\delta_{ik}\delta_{jm}+\delta_{im}\delta_{jk}
\right) \label{eq24}
\end{equation}
and $H$ is given by (\ref{eq22}) [cf. Eq. (\ref{eq1})], which is a
conserved functional. In the functional derivatives (\ref{eq23})
the 18 components $B_{km}$ and $E_{km}$ are treated as if they
were independent. This is due to the fact that the right-hand
sides of Eqs. (\ref{eq23}), restricted to the submanifold $N$
defined by the conditions $E_{ij}=E_{ji}$, $E_{ii}=0$,
$B_{ij}=B_{ji}$ and $B_{ii}=0$, are symmetric in the indices $i$
and $j$ and have vanishing trace; therefore the evolution curves
given by Eqs. (\ref{eq23}) are tangent to $N$.

Making use of the $D_{ijkm}$ given by Eq. (\ref{eq24}), a Poisson
bracket between any pair of functionals of the field $F$ and $G$
can be defined as
\begin{equation}
\left\{ F, G \right\}\equiv \int \left( \frac{\delta F}{\delta
E_{ij}}D_{ijkm}\frac{\delta G}{\delta B_{km}} - \frac{\delta
F}{\delta B_{ij}}D_{ijkm}\frac{\delta G}{\delta E_{km}}\right)dv =
\int \left( \frac{\delta F}{\delta E_{km}}\frac{\delta G}{\delta
B_{km}} - \frac{\delta F}{\delta B_{km}}\frac{\delta G}{\delta
E_{km}}\right)dv.\label{eq25}
\end{equation}
The bracket (\ref{eq25}) is antisymmetric and satisfies the Jacobi
identity due to the fact that the $D_{ijkm}$ are constants [1].
Hence the $D_{ijkm}$ define a hamiltonian structure. From Eq.
(\ref{eq25}) one finds that
\begin{equation}
\left\{E_{ij}(\textbf{r}', t),E_{km}(\textbf{r}'', t)
\right\}=0=\left\{B_{ij}(\textbf{r}', t),B_{km}(\textbf{r}'', t)
\right\} \label{eq26}
\end{equation}
and that
\begin{equation}
\left\{E_{ij}(\textbf{r}', t),B_{km}(\textbf{r}'', t)
\right\}=D_{ijkm}\delta(\textbf{r}'-\textbf{r}'')\label{eq27}
\end{equation}
which are consistent with Eqs. (\ref{eq20}) since $\partial_{i}
D_{ijkm} =0$. Furthermore, with respect to the Hamiltonian
structure given by $D_{ijkm}$, one can find easily that the
functionals
\begin{equation}
P_{k}=\int\frac{1}{2}\left(E_{ij}\partial_{k}B_{ij}-
B_{ij}\partial_{k}E_{ij}\right)dv=-\int B_{ij}\partial_{k}E_{ij}dv
\label{eq28}
\end{equation}
(see e.g. Ref.[2]) are the components of the momentum and they are
conserved by the invariance of $H$ in any direction.

If $F$ is any functional of the field that does not depend
explicitly on the time then Eqs. (\ref{eq23}) and (\ref{eq25})
give
\begin{equation}
\left\{ F, H \right\}= \int \left( \frac{\delta F}{\delta E_{ij}}
\dot{E}_{ij} - \frac{\delta F}{\delta B_{ij}}
\dot{B}_{ij}\right)dv = \dot{F} \label{eq29}
\end{equation}

Similarly, by using Eqs. (\ref{eq25}),(\ref{eq28}) and
(\ref{eq20}), and the fact that the right-hand sides of Eqs.
(\ref{eq21}) are symmetric in the indices $i$ and $j$ one obtains
\begin{equation}
\left\{ F, P_{k} \right\}= - \int \left( \frac{\delta F}{\delta
E_{ij}} \partial_{k}E_{ij} + \frac{\delta F}{\delta B_{ij}}
\partial_{k} B_{ij}\right)dv \label{eq30}
\end{equation}
which means that, with respect to the Hamiltonian structure
associated with the bracket (\ref{eq25}), the functional $P_{k}$,
is, in effect, the generator of the translations in the direction
of the axis $x^{k}$. It is in this sense that the $P_{k}$ are the
components of the momentum of the field.

In contrast with Eq. (\ref{eq28}), when $H$ is given by the analog
of the energy of the field, the functionals
\begin{equation}
P_{k}=\int \frac {\kappa}{c}\epsilon_{kim}E_{ij}B_{mj}dv
\label{eq31}
\end{equation}
are the components of the momentum of the field [3].

\section{Concluding remarks}
\,\,\,\ The Hamiltonian employed in this example is a conserved
functional of the field, however it is not know if there exist
additional conditions for a conserved functional to be a
Hamiltonian, with the corresponding Poisson bracket satisfying the
Jacobi identity (when the $D_{ab}$ are constants the Jacobi
identity is always satisfied, but in other cases one has to verify
that this identity is satisfied [1]). In the case of a mechanical
system with a finite number of degrees of freedom in classical
mechanics, any constant of motion can be used as Hamiltonian by
defining appropriately the symplectic structure of the phase space
(or, equivalently, the Poisson bracket) [4].

The example considered here shows a different form to the
traditional canonical formalism to write the evolution equations
for an infinite-dimensional system, in which sometimes there are
constraints. In the present case the $D_{ab}$ are constants
because $H$ depends on the $\partial_{i}\phi _{a}$ (and on the
$\phi _{a}$, of course), this is an advantage since one can obtain
the components of the momentum of the field immediately [2].

\section*{Acknowledgments}
The author is grateful to Professor G. F. Torres del Castillo for
stimulating discussions. This work was financially supported by
CONACYT (M\'exico).

\end{document}